\documentclass[sigconf]{acmart}
\AtBeginDocument{%
  \providecommand\BibTeX{{%
    \normalfont B\kern-0.5em{\scshape i\kern-0.25em b}\kern-0.8em\TeX}}}
\usepackage{graphicx,tabularx,subcaption}
\usepackage{booktabs}
\usepackage{multirow}
\usepackage[capitalise,nameinlink,compress]{cleveref}
\usepackage{colortbl}
\usepackage{xcolor}
\usepackage{comment}
\usepackage[english]{babel}
\usepackage{fontawesome5}
\usepackage{arydshln}
\usepackage{tikz}
\usepackage{enumitem}
\graphicspath{{images/}}
\usepackage{subcaption}
\usepackage{framed}

\newenvironment{frshaded*}{%
\MakeFramed {\advance\hsize-\width \FrameRestore}
}%
{\endMakeFramed}

\definecolor{main}{HTML}{b3dd9f}    
\definecolor{sub}{HTML}{EEF9E9}     

\definecolor{lightgray}{gray}{0.9}

\definecolor{greenShade}{RGB}{239, 255, 232}
\definecolor{greenFont}{RGB}{34, 139, 34}

\definecolor{pinkShade}{RGB}{254, 236, 240}
\definecolor{pinkFont}{RGB}{251, 29, 71}

\definecolor{lightBlueShade}{RGB}{217, 243, 255}
\definecolor{lightBlueFont}{RGB}{0, 122, 176}

\definecolor{peachShade}{RGB}{255, 231, 209}
\definecolor{peachFont}{RGB}{242, 115, 0}

\definecolor{blueShade}{RGB}{231, 239, 253}
\definecolor{blueFont}{RGB}{19, 99, 223}

\definecolor{purpleShade}{RGB}{241, 232, 248}
\definecolor{purpleFont}{RGB}{112, 48, 160}

\definecolor{yellowShade}{RGB}{255, 253, 235}
\definecolor{yellowFont}{RGB}{204, 160, 29}

\definecolor{orangeShade}{RGB}{252, 230, 208}
\definecolor{orangeFont}{RGB}{249, 111, 7}

\definecolor{beigeShade}{RGB}{246, 237, 226}
\definecolor{brownFont}{RGB}{116, 81, 45}

\definecolor{softYellowShade}{RGB}{255, 255, 221}
\definecolor{darkYellowFont}{RGB}{255, 212, 0}

\copyrightyear{2025} 
\acmYear{2025} 
\setcopyright{acmlicensed}
\acmConference[UMAP '25]{33rd ACM Conference on User Modeling, Adaptation and Personalization}{June 16--19, 2025}{New York City, NY, USA}
\acmBooktitle{33rd ACM Conference on User Modeling, Adaptation and Personalization (UMAP '25), June 16--19, 2025, New York City, NY, USA}
\acmDOI{10.1145/3699682.3728321}
\acmISBN{979-8-4007-1313-2/2025/06}

\begin{document}


\title[Benefits and Challenges of Agent-Augmented Counterfactual Explanations for Non-Expert Users]{``Show Me How'': Benefits and Challenges of Agent-Augmented Counterfactual Explanations for Non-Expert Users}

\author{Aditya Bhattacharya}
\orcid{0000-0003-2740-039X}
\email{aditya.bhattacharya@kuleuven.be}
\affiliation{%
  \institution{KU Leuven}
  \city{Leuven}
  \country{Belgium}
}
\authornote{Both authors contributed equally to this research.}

\author{Tim Vanherwegen}
\orcid{0009-0009-1928-8271}
\email{tim.vanherwegen@student.kuleuven.be}
\affiliation{%
  \institution{KU Leuven}
  \city{Leuven}
  \country{Belgium}
}
\authornotemark[1]

\author{Katrien Verbert}
\orcid{0000-0001-6699-7710}
\email{katrien.verbert@kuleuven.be}
\affiliation{%
  \institution{KU Leuven}
  \city{Leuven}
  \country{Belgium}
}

\renewcommand{\shortauthors}{Bhattacharya and Vanherwegen, et al.}

\begin{abstract}

Counterfactual explanations offer actionable insights by illustrating how changes to inputs can lead to different outcomes. However, these explanations often suffer from ambiguity and impracticality, limiting their utility for non-expert users with limited AI knowledge. Augmenting counterfactual explanations with Large Language Models (LLMs) has been proposed as a solution, but little research has examined their benefits and challenges for non-experts. To address this gap, we developed a healthcare-focused system that leverages conversational AI agents to enhance counterfactual explanations, offering clear, actionable recommendations to help patients at high risk of cardiovascular disease (CVD) reduce their risk. Evaluated through a mixed-methods study with 34 participants, our findings highlight the effectiveness of agent-augmented counterfactuals in improving actionable recommendations. Results further indicate that users with prior experience using conversational AI demonstrated greater effectiveness in utilising these explanations compared to novices. Furthermore, this paper introduces a set of generic guidelines for creating augmented counterfactual explanations, incorporating safeguards to mitigate common LLM pitfalls, such as hallucinations, and ensuring the explanations are both actionable and contextually relevant for non-expert users.
\end{abstract}





\maketitle

\section{Introduction}
Explainable artificial intelligence (XAI) methods are crucial for interpreting ``black-box'' machine learning (ML) and artificial intelligence (AI) models \cite{adadi2018peeking, BhattacharyaXAI2022, BhattacharyaCHIDC}. Among the diverse XAI approaches, counterfactual explanations stand out as example-based methods that offer \textit{actionable recourse} for end users \cite{ustun2019}, providing recommendations for minimal changes needed to achieve a favourable prediction during informed decision making \cite{adadi2018peeking, BhattacharyaXAI2022}. Despite the popularity of counterfactual explanations, they face significant limitations in practical applications. Their practical plausibility is often restricted, as the suggested changes may not be feasible or actionable in real-world scenarios \cite{keane2021bettercounterfactualexplanationskey, Bhattacharya2023, spreitzer2022evaluating}. 

Moreover, these explanations heavily depend on the training data, often overlooking contextual knowledge, feature interdependence, and a broader knowledge base, which can result in impractical recommendations \cite{keane2021bettercounterfactualexplanationskey}. For instance, for a diabetes prediction use case, counterfactual algorithms may recommend an aged individual (suppose 80 years old) with existing heart conditions for intensive outdoor running to reduce the risk of diabetes. Therefore, instead of improving their medical conditions, such recommendations can lead to severe adverse effects. Additionally, counterfactual explanations may generate contradictory suggestions due to the lack of contextual and in-depth real-world knowledge, further complicating their utility and reliability for non-expert users with limited AI knowledge \cite{BhattacharyaXAI2022, Bhattacharya2023}. 

To address these limitations, prior research has considered using Large Language Models (LLMs) to refine and tailor counterfactual explanations for end-users \cite{fredes2024usingllmsexplainingsets, AmritaBhattacharjee2024, ijcai2018p836}. By leveraging extensive contextual information and practical knowledge beyond the training data, LLMs can generate more relevant and context-aware counterfactuals. However, given the known pitfalls of LLMs, such as hallucinations and biased outputs, prior work has emphasised the need for extensive user studies involving non-expert users with limited AI knowledge to better understand the benefits and challenges of LLMs for generating explanations \cite{fredes2024usingllmsexplainingsets, kunz-kuhlmann-2024-properties, schiller2024humanfactordetectingerrors, tjuatja-etal-2024-llms}.

Our work examines the advantages and limitations of augmenting counterfactual explanations with LLM-based conversational AI agents with non-experts with varying AI proficiency
levels. To explore this, we developed a healthcare-focused conversational system that enables non-experts with limited AI knowledge (such as patients) to interact with a cardiovascular disease prediction model to receive actionable recourse. The system was designed using a user-centric approach, beginning with an exploratory study involving four participants to identify initial user requirements and application features. This was followed by a mixed-methods study with 34 participants to evaluate the system’s effectiveness.

Understanding how different user groups interact with conversational AI is essential for designing intuitive and trustworthy systems \cite{Slack2023, lakkaraju2022rethinking}. Prior work suggests that user familiarity with AI-driven tools can shape their expectations, trust, and ability to interpret AI-generated explanations \cite{Schreuter2021, bove_contextualization_2022}. Therefore, our study aimed to investigate whether prior experience with conversational agents, such as chatbots, influenced users’ interaction patterns. To examine this, we categorised participants into two groups: (1) \textit{novice} users with little to no experience using conversational AI applications and (2) \textit{informed} users who had prior exposure to such tools. We particularly delved into the following research questions:

\begin{description}[topsep=0pt, itemsep=0pt, left=-0.01cm]
    \item[RQ1.] How do novice and informed end users utilise agent-augmented counterfactuals to achieve actionable recourse?

    \item[RQ2.] How do conversational agents impact the understanding and trust of novice and informed end users?

    \item[RQ3.] How does perceived taskload differ between novice and informed users when using the chatbot application?
\end{description}

\noindent In summary, our work provides the following key contributions:
\begin{enumerate}[topsep=0pt, itemsep=0pt, left=-0.01cm]
    \item  \textbf{Theoretical Contribution}: We present a set of generic guidelines for augmenting counterfactual explanations with conversational AI agents. These guidelines aim to mitigate the known limitations of counterfactual generation algorithms and LLMs for more relevant and context-aware explanations.
    \item  \textbf{Artifact Contribution}: We instantiated these guidelines into a healthcare chatbot application that allows end users to interact with a cardiovascular disease prediction model to guide them in obtaining their desired predictions. The source code, design, and architecture of this system are open-sourced on \href{https://github.com/adib0073/ShowMeHow/}{GitHub}.
    \item  \textbf{Empirical Contribution}: Our work empirically examines the benefits and drawbacks of agent-augmented counterfactuals through user studies involving novice and informed non-experts.
\end{enumerate}

\section{Background and related work}\label{sec_lit_study}
\subsection{Counterfactual Explanations}
Counterfactual explanations assist users by presenting alternative instances, or counterfactuals, that could lead to a different outcome \cite{BhattacharyaXAI2022}. This explainable AI (XAI) method is particularly valuable for explaining AI-based decision support systems that negatively impact individuals \cite{singh2021directive}. For instance, if an AI-based hiring system rejects a qualified candidate, it should at least explain the steps the applicant can take to improve their chances of being selected in the future. Counterfactual explanations can facilitate this by not only clarifying \textit{why} the model produced a particular decision but also guiding users on \textit{how} they can alter their circumstances to achieve a more favourable outcome, if feasible. This act of providing recommendations to achieve a desired outcome through counterfactual algorithms is also referred to as \textit{actionable recourse} \cite{ustun2019}.

Wachter et al.~\cite{wachter2017counterfactual} identify three key purposes for counterfactual explanations: i) explaining why a particular decision was made, ii) giving users grounds to contest the decision, and iii) offering actionable steps to reverse the outcome. While Wachter et al. argue that counterfactual explanations can meet all three objectives, other researchers have noted that generating practically feasible counterfactuals that satisfy these conditions is challenging due to the lack of contextual knowledge and myopic nature of counterfactual generation algorithms \cite{russell2019efficient, singh2021directive}. Also, concerns about the practical feasibility of counterfactual examples highlight the need to vet them thoroughly to ensure that the recourses they offers are meaningful and non-discriminatory for different users \cite{mahajan2019}.

To address the limitations of counterfactual generation algorithms, augmenting counterfactuals with LLM-based AI agents has been proposed as a potential solution \cite{fredes2024usingllmsexplainingsets, AmritaBhattacharjee2024, ijcai2018p836}. This approach offers two key benefits: i) leveraging the broader knowledge base of LLMs to refine counterfactuals, ensuring only feasible actions that are neither contradictory nor confusing are suggested to users, and ii) facilitating dialogue-based interactions that help users better understand the recommendations and allow them to provide iterative feedback for fine-tuned guidance based on their specific needs. Our work investigates the main benefits and challenges of such augmented counterfactuals from the perspective of non-experts.

\subsection{Conversational XAI using AI Agents}
Prior research has highlighted the value of conversational explanations, delivered through free-form conversations, in enhancing user understanding of static explanations generated by XAI methods \cite{Slack2023, lakkaraju2022rethinking, zhang2024iaskfollowupquestion, luo2023providingpersonalizedexplanationsconversational, Miller2017}. These explanations leverage natural language dialogue to deliver dynamic, personalised responses tailored to the user’s background, needs, and preferences \cite{shen2023convxaideliveringheterogeneousai, Slack2023, lakkaraju2022rethinking, zhang2024iaskfollowupquestion}. Recent advancements in LLM-based AI agents have brought significant attention to context-aware conversational explanations, highlighting their potential for generating actionable insights \cite{nguyen2023black}.

AI agents are autonomous systems powered by LLMs designed to simulate human-like conversations \cite{Marini2023CA}. While LLMs primarily generate text, they lack the inherent ability to execute direct actions. However, when integrated into AI agents, LLMs function as reasoning engines that identify appropriate actions and the required inputs for those actions. The outcomes are then fed back into the agent, enabling it to evaluate whether further steps are necessary or if the interaction can be concluded effectively.



Despite the benefits of conversational AI agents, they are prone towards \textit{hallucination}. Hallucination in the context of LLMs is defined as the act of generating content that is factually incorrect, inconsistent or completely irrelevant considering the real-world facts or user inputs \cite{huang2023surveyhallucinationlargelanguage}. The two most effective approaches proposed in the literature to mitigate the hallucination are: (1) \textsc{Prompt engineering} and (2) \textsc{Fine-tuning} \cite{huang2023surveyhallucinationlargelanguage, li-etal-2024-dawn, xu2024hallucinationinevitableinnatelimitation}. Prompt engineering is the process of crafting effective instructions (or \textit{prompts}) to guide the LLM in generating desired outputs. Whereas fine-tuning is the process of customising a pre-trained LLM to perform specific tasks by training it on a smaller and more relevant dataset. In this paper, we present general guidelines for enhancing counterfactual explanations through conversational AI agents that offer contextual knowledge, mitigate hallucinations, and refine counterfactual examples for greater clarity and relevance.

\section{Guidelines for Generating Agent-Augmented Counterfactuals}\label{sec_framework}
This section presents our general guidelines for creating agent-augmented counterfactual explanations. This process can be subdivided into two parts: the first part focuses on steering the conversational agent to increase contextual knowledge and avoid common LLM pitfalls. The second part enriches counterfactual explanations, ensuring they are practically relevant and offer meaningful recommendations to users for achieving actionable recourse.

\emph{ \textbf{Methodology for Guideline Formulation}}: We conceptualised these guidelines for agent-augmented counterfactuals through an extensive literature review, synthesising insights from Explainable AI, LLMs, and AI agents to ensure comprehensive coverage. Our structured approach began with identifying the limitations of counterfactual generation algorithms in producing actionable recommendations for non-expert users. We then explored mitigation strategies using LLMs, followed by examining the strengths and weaknesses of LLM-generated conversational explanations. Lastly, we analysed research on conversational AI agents to enhance explanation methods. These guidelines were iteratively refined based on the feedback from our user studies.

\subsection{Steering the Conversational Agent}

\subsubsection{\textbf{Context-Fusion Prompting}} To impart relevant contextual knowledge into LLMs used in AI Agents, we echo the thoughts of Wang et al. \cite{wang-etal-2024-browse} for the necessity of prior context fusion of LLMs. This contextual information should be incorporated through initial prompts to fine-tune the agent's responses. We recommend creating a comprehensive data dictionary of the training dataset used by the prediction model to capture the contextual knowledge. This document should detail the predictor variables, including descriptions, permissible value ranges, units of measurement, and practical implications of encoded variables. To further reduce hallucinations and irrelevant responses, we suggest including local inference data, i.e., the specific information that is going to be used for generating the predictions. This local information could be particularly useful if the user wants to conduct a \textit{what-if analysis} \cite{BhattacharyaXAI2022, BhattacharyaUMAP24Demo, Grafberger2022whatif} through multiple dialogues.
\subsubsection{ \textbf{Tools for Moderation Check}} LLMs are susceptible to common issues such as hallucinations, harmful queries, and even malicious attempts by users to manipulate the model’s behaviour through \textit{prompt injections} \cite{liu2024promptinjectionattackllmintegrated, feldman2023trappingllmhallucinationsusing, kumar2024watchlanguageinvestigatingcontent}. To mitigate these risks, we recommend equipping the AI agent with explicit \textit{tools} (i.e., utility functions) to validate user queries and flag any violations of moderation guidelines. This moderation check should be performed for every user input before passing the query to the LLM. In cases where a violation is detected, the agent should generate a standard response, asking the user to avoid queries containing harmful content or attempts to manipulate the LLM’s behaviour.
\subsubsection{ \textbf{Tools for Counterfactual Generations}} To generate recommendations in the appropriate format from counterfactual generation algorithms, the agents should have access to tools that apply a trained ML model to inference data for outcome prediction, followed by applying counterfactual algorithms to generate counterfactual instances. Generally, counterfactual generation algorithms produce multiple counterfactual instances, making it challenging to select the most relevant one. However, this issue can be mitigated by leveraging conversational agents, which can be guided through follow-up dialogues to choose the most relevant and useful recommendations for the user. Moreover, we suggest adding prompts using Chain-of-Thought (CoT) prompting \cite{wang-etal-2023-towards} and ReAct prompting \cite{yao2023reactsynergizingreasoningacting} guidelines to generate causal reasoning for the recommendations and further justify why these actions are recommended to the user.

\subsubsection{ \textbf{Tools for Self Reflection}} Once the counterfactual-based recommendations are generated, we suggest incorporating additional tools to validate their feasibility. Specifically, we recommend using the \textit{LLM-as-a-Judge} approach \cite{zheng2023judgingllmasajudgemtbenchchatbot} to assess the practicality and actionability of the recommendations. When setting up the validation prompt, we advise reintroducing the local inference data to cross-verify whether the recommendations are appropriate for the specific instance. Based on the final evaluation of the LLM-as-a-Judge approach, the most appropriate recommendation will be shared with the user.

\subsection{Enriching Counterfactual Explanations}

\subsubsection{\textbf{Generate Counterfactuals for Actionable Features}} To prevent recommending changes to factors that are not practically feasible to modify (i.e., non-actionable features), this component emphasises including only actionable features when generating counterfactual examples similar to prior work \cite{Bhattacharya2023, bhattacharya2024exmos, bhattacharya2023_technical_report}. The tool used by the conversational agent for generating counterfactual examples should ensure that the algorithms have access only to predictor variables that are actionable, thereby producing more practical and relevant recommendations.

\subsubsection{\textbf{Guardrails for Counterfactuals}} Generally, counterfactual generation algorithms tend to overlook the association between predictor variables, treating each variable as independent to each other \cite{keane2021bettercounterfactualexplanationskey}. As a result, they may suggest counter-intuitive actions. For instance, consider a diabetes prediction model that identifies an over-weight, young patient as high-risk based on multiple health measures, with physical activity levels being one of them. A counterfactual algorithm might recommend reducing physical activity to lower risk for the young overweight patient, which would be illogical in practice. A medical expert would never advise reducing physical activity for such a patient unless specific health concerns exist. To prevent such counter-intuitive recommendations, we suggest implementing guardrails through a rule-based algorithm to post-process the generated counterfactual instance, ensuring they align with real-world expectations.

\subsubsection{\textbf{Supplement Counterfactuals with Data-Centric Explanations}} To further enrich counterfactual explanations, we recommend supplementing them with visually directive data-centric explanations as implemented by Bhattacharya et al. \cite{Bhattacharya2023}. We suggest adding interactive data-centric explanations that provide a local explanation with a global overview so that users can better understand the counterfactual recommendations. These data-centric explanations would further help them explore how the model's behaviour changes if the underlying data changes. Users can additionally perform data-centric what-if analysis \cite{Grafberger2022whatif} to provide feedback to the conversational agent for further fine-tuning the recommended actions.

\section{Chatbot Application}
\label{sec_app}

\subsection{Usage Scenario}
Building on the guidelines for agent-augmented counterfactuals discussed in \Cref{sec_framework}, we developed a chatbot application tailored to monitoring cardiovascular disease (CVD) risk. The system integrates an ML model that predicts CVD risk based on patient medical records, helping users assess and understand their heart disease risk. It supports users by highlighting critical health factors requiring immediate attention and allowing feedback to refine recommendations. Additionally, the chatbot enables users to justify key risk factors, explore strategies for improving their condition, and evaluate the impact of specific lifestyle changes.

\begin{figure*}
\centering
\includegraphics[width=0.9\linewidth]{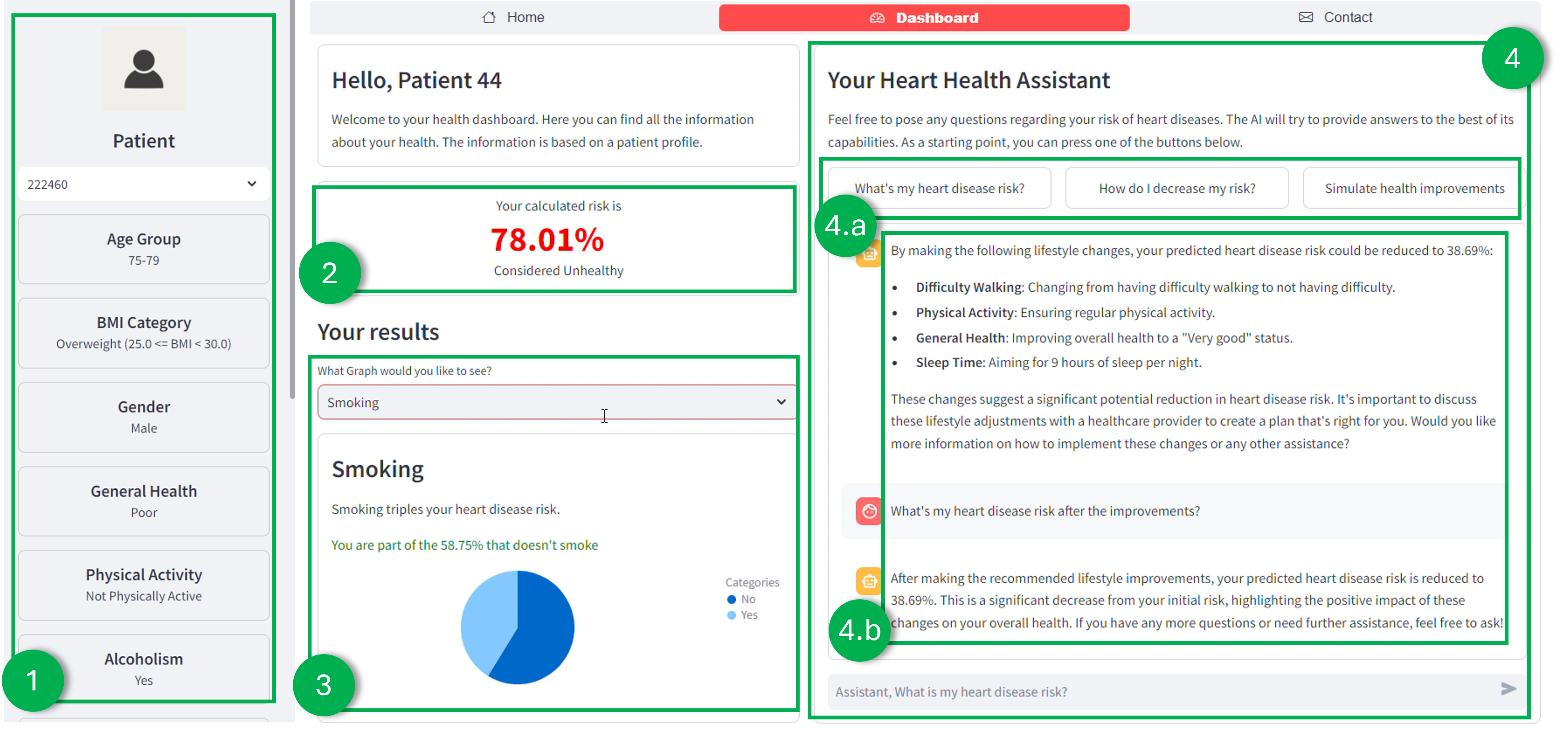}
\caption{Screenshot of our chatbot application illustrating the UI components described in \Cref{sec_ui_components}: (1) Patient Information (2) Risk Status (3) Visual Explanations (4) Chatbot Assistant (4.a) Ice-breaker Questions (4.b) Agent-Augmented Counterfactuals }
\Description[Application Screenshot]{Screenshot of our chatbot application illustrating the UI components described in \Cref{sec_ui_components}: (1) Patient Information (2) Risk Status (3) Visual Explanations (4) Chatbot Assistant (4.a) Ice-breaker Questions (4.b) Agent-Augmented Counterfactuals}
\label{fig:ui_components}
\end{figure*}

\subsection{Application Implementation}
\textbf{\textit{User-centric design approach}}: The application was developed following a user-centered design process \cite{UCD2014}. We began by creating a low-fidelity prototype based on the design guidelines established by Yang and Aurisicchio \cite{yangdesigningchatbot}. This prototype was implemented as a click-through interface using Figma \cite{figmaUrl} and served as the foundation for an exploratory user study. The study, conducted through think-aloud sessions with four participants, provided valuable insights into user needs and interaction patterns. Each think-aloud session lasted approximately 30 minutes. The participants were recruited through social media platforms for voluntary pro-bono participation. They ranged in age from 22 to 76, comprising two males and two females. This study also included showing the visual counterfactual explanation design proposed by Bhattacharya et al. \cite{Bhattacharya2023}, in which counterfactual explanations are presented as actionable recommendations. This approach did not involve augmenting the counterfactual examples using an LLM but served as a baseline for identifying key limitations of non-augmented counterfactual explanations. 
These findings informed the development of UI components that support key user requirements for achieving actionable recourse, as mentioned in the following part. The refined high-fidelity prototype is further detailed in \Cref{sec_ui_components}.

\noindent \textbf{\textit{User requirements}}: 
The exploratory study resulted in the formulation of the following key user requirements. To fulfil these user requirements, we then designed and developed the UI components described in \Cref{sec_ui_components}.
\begin{enumerate}[start=1,label={\textbf{ \arabic*.}}, left=0cm]
    \item \textbf{Guided conversation starter}: In our exploratory study, we observed that participants had difficulty initiating conversations with the chatbot. They suggested providing \textit{``ice-breaker question''} to better understand the chatbot's capabilities and purpose. This aligns with prior research emphasising the importance of guided starter questions to ease users into the conversation \cite{Slack2023, freed2021conversational, zhang2024navigatinguserexperiencechatgptbased}. As a result, we included ice-breaker questions as a key feature in our high-fidelity prototype.

    \item \textbf{Highlighting factors that need immediate attention}: Participants emphasised the importance of prioritising recommended actions by highlighting important factors that contribute most to elevated CVD risk. In response, we refined the design of our visual data-centric explanations to highlight factors requiring immediate attention, drawing inspiration from Bhattacharya et al.'s approach \cite{Bhattacharya2023}.

    \item \textbf{Ability to provide feedback}: Participants underscored the value of providing feedback to further refine recommended actions. In response, our high-fidelity prototype included a feature that allowed users to engage with follow-up questions, enabling fine-tuning of recommendations generated by counterfactual algorithms to better align with individual user needs.

\end{enumerate}

\noindent \textbf{\textit{Chatbot application}}: Following our generic guidelines for creating agent-augmented counterfactuals, we developed a high-fidelity application that facilitates non-experts in achieving actionable recourse. This chatbot application was developed using Streamlit \cite{streamlit}, a Python framework used for developing web-based applications. We used the \textsc{gpt-4-turbo} model from OpenAI \cite{openai_online} as the LLM in our conversational agent. Moreover, we used LangChain \cite{langchain_online}, an open-sourced framework designed to support the development of robust LLM applications. 
Additionally, for steering the conversational agent, we developed custom tools for moderation checks using OpenAI's moderation API \cite{openai_moderation_guide}, for detecting prompt injection attacks following the prompting guidelines from AWS Prescriptive Guidance \cite{AWS2024prompt} and followed the Chain-Of-Verification prompt engineering to minimise hallucination  \cite{dhuliawala2023chainofverificationreduceshallucinationlarge}.

\noindent \textbf{\textit{Prediction model}}: Our application included a deep neural network model for predicting the likelihood of cardiovascular disease (CVD) from patient's medical records. This prediction model had an accuracy of 91.4\%. While our proposed guidelines and the chatbot application are model-agnostic as they do not depend on the type of prediction algorithm used, we selected a deep neural network model for its higher accuracy and minimal over-fitting effect on the training dataset. The trained model was made available as a tool to the conversational agent for generating real-time predictions and counterfactual examples based on user queries.

\noindent \textbf{\textit{Dataset}}: The model was trained on an open-sourced CVD prediction dataset \cite{kaggle_heart_disease_dataset}. This dataset was compiled by the Centers for Disease Control and Prevention (CDC) and is a critical component of the Behavioral Risk Factor Surveillance System (BRFSS) \cite{brfss}. It consists of 319,796 patient records and 18 columns (17 predictor variables and one target variable). The dataset includes information about several health factors relevant for predicting CVD. The comprehensive dataset description was provided to the conversational agent through prompt engineering before initialising the conversation with users to add more contextual knowledge.  

\noindent \textbf{\textit{Counterfactual generation}}:  To allow users to explore what-if scenarios for the predicted CVD risk, we integrated an on-demand counterfactual explanation generation tool into the conversational agent. The counterfactuals were generated using the DiCE Python framework \cite{dice_framework}. Additionally, considering our guidelines, the counterfactuals were generated only for actionable variables (i.e., health factors that could be modified by patients like BMI or glucose levels). Furthermore, the chatbot included visually directive data-centric explanations \cite{Bhattacharya2023}, enabling users to interactively modify input values and perform what-if analyses.

\subsection{User Interface Components}\label{sec_ui_components}
This section describes the following UI components of our chatbot application (as illustrated in \Cref{fig:ui_components}):

\begin{enumerate}[start=1,label={\textbf{ \arabic*.}}, topsep=0pt, itemsep=0pt, left=-0.01cm]

 \item \textbf{Patient Information}: This UI component presents the health measures of a selected patient, which are utilised by the trained prediction model to estimate the probability of CVD risk. These details are shared with the agent to integrate local contextual knowledge about the patient and displayed to the patient to enhance their awareness of critical health factors.

 \item \textbf{Risk Status}: This UI component highlights the predicted CVD risk score to improve patients' awareness of their risk status. A score closer to 100 indicates a high risk of CVD, while a score below 50 signifies a low risk.

 \item \textbf{Visual Explanations}: This UI component presents a clear overview of a patient's health metrics using interactive data distribution visualisations (similar to \cite{Bhattacharya2023}). The system presents optimal health measures for achieving a low CVD risk and illustrates the discrepancies between the patient’s current metrics and the recommended ranges through visual aids. Users can adjust the predictor variable values to see the effect on the overall risk score. Variables with significant deviations from ideal values for reducing CVD risk are flagged with warning messages to draw users' immediate attention.

 \item \textbf{Chatbot Assistant}: The chatbot component facilitates user interaction with the backend AI agent. Based on feedback from the exploratory study, we incorporated suggested questions as guided conversation starters or \textit{ice-breaker questions}. These questions help users understand their current CVD risk without delving into visual explanations and offer insights into reducing risk through augmented counterfactuals. Each recommendation to lower risk is accompanied by detailed justifications to support causal reasoning. The chatbot also enables what-if analysis, allowing users to propose alternate scenarios through dialogue and receive explanations on how these changes affect their predicted risk scores. Furthermore, if a suggested action is deemed impractical, the agent can adapt its recommendations, providing alternative counterfactuals tailored to the user's input.
\end{enumerate}

\section{Final Evaluation}\label{sec_study}

\subsection{Study Setup}
The final evaluation of our chatbot application was conducted through a mixed-methods user study involving 34 participants. The study protocols were approved by the ethical committee of \anon{KU Leuven} (approval number: \anon{G-2024-7704}). This study was conducted online through Google Forms. On average, each participant took around 45 minutes to complete their participation. 

\subsection{Participants}
Participants for this study were voluntarily recruited through social media platforms, primarily from heart disease discussion groups on Facebook and Reddit, and a local Pilates studio in \anon{Leuven, Belgium}. Eligibility was limited to adults (18+ years) with minimal or no experience using AI applications. The study included 34 participants, comprising 14 novice users with limited awareness of conversational AI and 20 informed users with prior experience using conversational AI applications but no technical AI knowledge. Participants ranged in age from 19 to 57 years ($mean: 28, SD: 9.8$), with 14 identifying themselves as male and 20 as female. Furthermore, we selected participants with prior experience and knowledge related to cardiovascular diseases and their associated risks from a patient’s perspective.





\subsection{Evaluation Measures}
For each of the following evaluation measures, we collected user perspectives through a combination of quantitative data and open-ended qualitative questions. The complete set of study questionnaires is provided in the supplementary material\footnote{Supplementary Material: \url{https://github.com/adib0073/ShowMeHow/raw/refs/heads/main/supplementary.zip}}. 

\emph{Perceived Actionability of the Agent}: Inspired by the work of Shoemaker et al. \cite{shoemaker2014pemat}, we define perceived actionability as the extent to which users believe that the information provided by the agent enables them to identify clear, feasible actions they can take to alter the decision of a prediction model. We assessed perceived actionability using both objective scores and subjective scores. Following the approach of Bhattacharya et al. \cite{Bhattacharya2023}, the objective scores were measured using task-based questions for achieving actionable recourse and the subjective scores were measured using 5-point Likert scale questions. 
Building on prior works \cite{Lim2009, PearlMackenzie2018, Bhattacharya2023}, we designed three task-based questions for our objective evaluation: \textit{Justification Task (T1)}, \textit{How-To Task (T2)}, and \textit{What-If Task (T3)}. For the justification task, participants interacted with the agent to identify the primary justification for a predicted CVD risk and determine which health factors contributed to the risk scores. In the how-to task, participants explored ways to improve a sample patient’s risk using the system. Finally, the what-if task required them to interact with the system to examine the impact of specific changes to actionable variables (e.g., reducing alcohol consumption) on the patient’s risk.

\emph{Perceived Understandability of the Augmented Counterfactuals}: Drawing on Hoffman et al.'s definition of perceived understandability of explanations \cite{hoffman2019metrics}, we define the perceived understandability of augmented counterfactuals as participants' confidence in comprehending the rationale behind the recommendations, knowing how to apply them effectively, and predicting their potential impact on decision-making, without requiring detailed knowledge of the underlying algorithms. To measure this, we adopted Hoffman et al.'s questionnaire on perceived understandability \cite{hoffman2019metrics}, using a 5-point Likert scale.

\emph{Perceived Trust in AI Agents}: Inspired by the definition of perceived trust in automated systems by Jian et al.\cite{Jian2020_trust_scale}, we define perceived trust as the user's confidence in the reliability, competence, and integrity of the agents when providing accurate, and relevant recommendation for achieving actionable recourse. This metric was also recorded on a 5-point Likert scale.

\emph{Perceived Taskload of the Application}: Perceived taskload refers to participants' subjective assessment of the mental, physical, and temporal demands experienced while interacting with the system, including the effort required to understand, process, and respond to the agent's recommendations. We used the NASA-TLX questionnaire to assess the perceived taskload of the chatbot application similar to prior researchers \cite{bhattacharya2024exmos, kulesza_explanatory_2010}. 

\emph{System Interaction Data}: In addition to the other evaluation measures, the system passively collected interaction data as participants engaged with the application. This data included the questions posed to the conversational AI, their entire conversation history and the interaction time spent by them on each UI component.

\subsection{Study Procedure}

Participants were first briefed on the study’s objectives, roles, and responsibilities and provided informed consent in accordance with ethical guidelines before submitting their demographic information. They then watched a detailed tutorial video and explored the application’s features through direct interaction. To assess objective actionability, participants completed three tasks (\textit{justification}, \textit{how-to}, and \textit{what-if} tasks) using examples from the test dataset of the trained ML model. Assuming the role of patients, they engaged with the system, allowing us to analyse their interactions and gather feedback relevant to our research questions. After completing the tasks, they evaluated subjective actionability, perceived understandability of augmented counterfactuals, trust in the agent, and overall perceived taskload.

\subsection{Data Analysis}
To explore whether users with varying levels of AI proficiency interacted differently with the system, we compared the responses of novice users with those of informed users during their interactions with our chatbot applications.  Since our data violated the normality assumptions \cite{mccrum-gardner_which_2008}, Mann-Whitney U-tests were conducted to observe if the differences between these two groups were statistically significant. 
Moreover, we performed thematic analysis using Braun and Clarke's method \cite{BraunClarkTA} for analysing the qualitative data collected from our study.

\section{Results}
\subsection{How do novice and informed end users utilise agent-augmented counterfactuals to achieve actionable recourse? (RQ1)}

Across the three task-based questions designed to achieve actionable recourse, 28 participants (82.3\%) successfully completed the \textit{justification task (T1)}, with an average completion time of 2 minutes. For the \textit{how-to task (T2)}, 32 participants (94.1\%) successfully completed it in approximately 3 minutes. Similarly, 30 participants (88.2\%) completed the \textit{what-if task (T3)}, taking about 2 minutes on average. These results demonstrate the potential of using augmented counterfactual explanations to facilitate actionable recourse in a relatively short time.

\begin{figure}[h]
\centering
\includegraphics[width=0.99\linewidth]{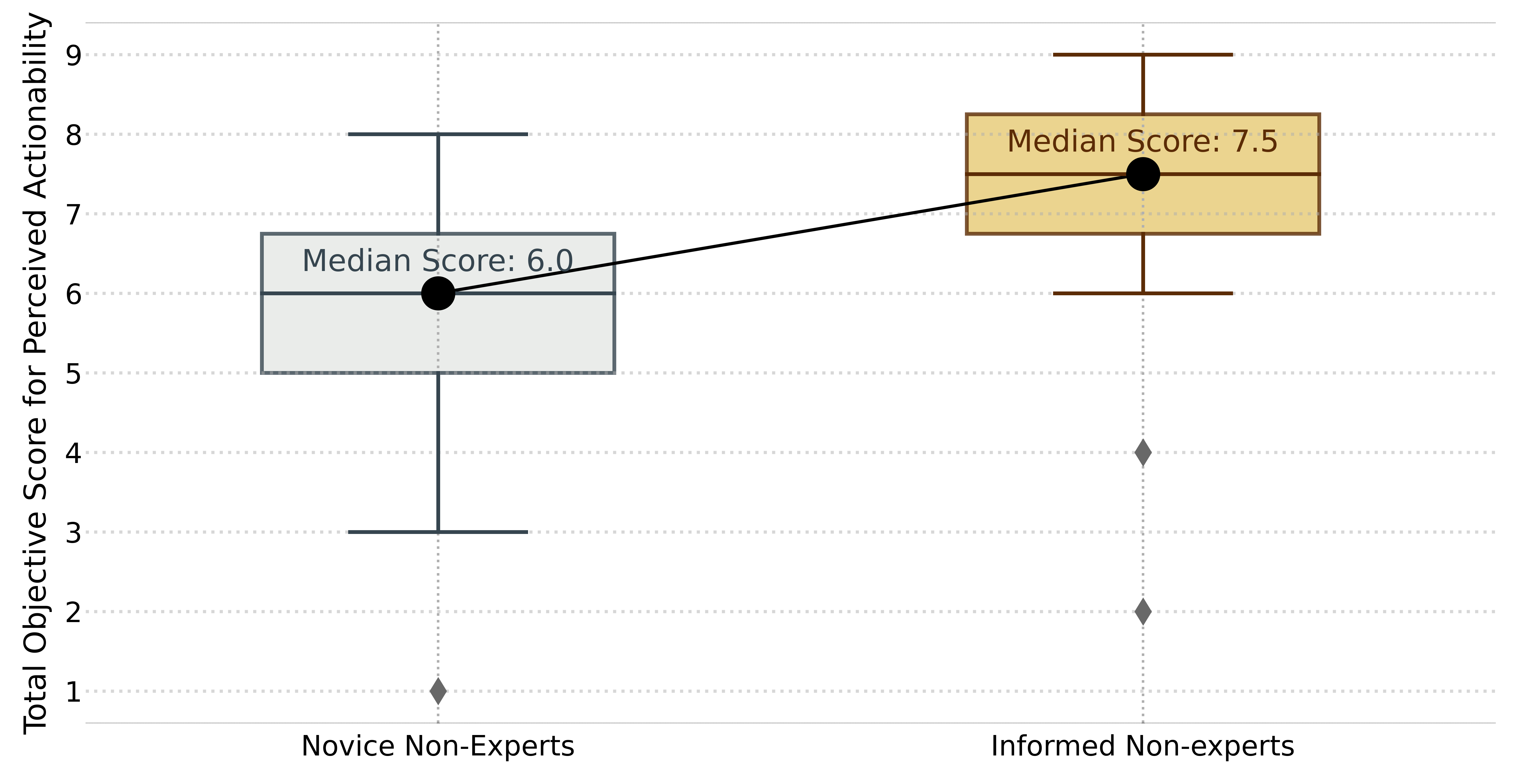}
\caption{Plots showing the difference in objective scores for perceived actionability between novice and informed users.}
\Description[Change in actionability task scores ]{Box plots showing the difference in scores for the task-based questions between novice and informed end users.}
\label{fig:task_scores}
\end{figure}

\begin{figure}[h]
\centering
\includegraphics[width=0.99\linewidth]{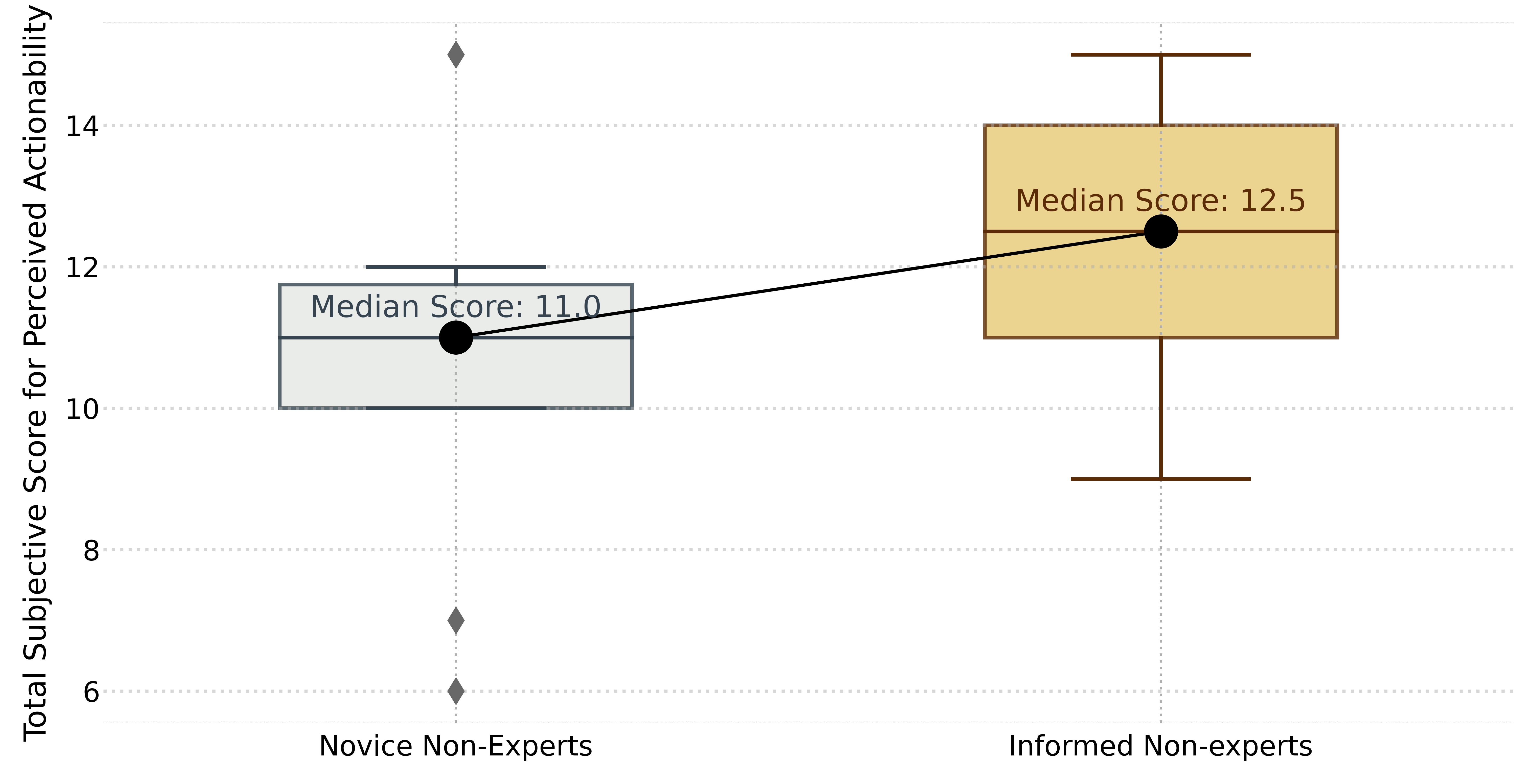}
\caption{Difference in subjective scores for perceived actionability between novice users and informed users.}
\Description[Change in actionability task scores ]{Box plots showing the difference in scores for subjective measurement of perceived actionability between Novice end users and Informed end users.}
\label{fig:subjective_actionability}
\end{figure}

Using Mann-Whitney U-test, we observed that \textit{informed users} were significantly better at answering the given task-based question than the novice users ($U=63.0, p=.006 $). On average, they achieved a higher score by approximately 19\%. Thus, our objective evaluation of perceived actionability indicates that informed users are better than the novice group in using augmented counterfactuals. \Cref{fig:task_scores} presents a box plot showing the difference in total objective score for perceived actionability between the two user groups. However, despite the informed users showing an increase in the subjective measure of perceived actionability than the novice users (illustrated in \Cref{fig:subjective_actionability}
), this difference was not statistically significant using a Mann-Whitney U-test ($U=91.5, p=.087$).

\begin{figure}[h]
\centering
\includegraphics[width=0.99\linewidth]{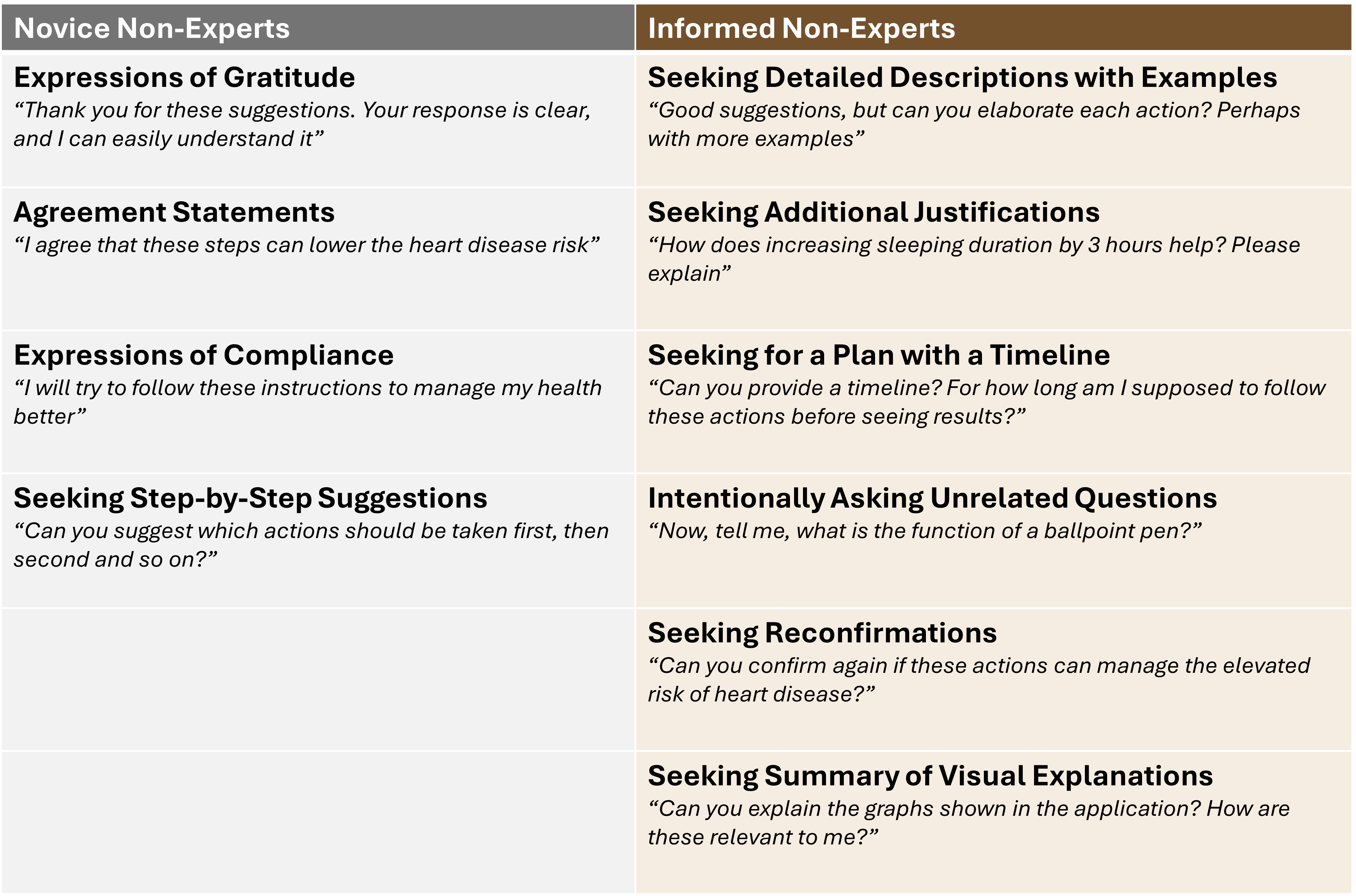}
\caption{This figure lists different utterance types for novice and informed users obtained using thematic analysis, along with example queries for as entered by the users.}
\Description[Types of utterances]{This figure lists out the different types of utterances from novice and informed users other than the task-based questions. For each utterance, we have provided an example query entered by a user.}
\label{fig:summary_conv_history}
\end{figure}

Thematic analysis of participants' qualitative responses and conversation histories helped us understand why the informed users perceived greater actionability of augmented counterfactuals than novice users. \Cref{fig:summary_conv_history} shows a list of utterance types from both groups (excluding task-based questions), which helps us investigate how each user group interacted with the chatbot.

(1) \emph{\textbf{In-depth follow-up questions from the informed users}}: From the qualitative data captured, we observed that informed users generally asked more follow-up questions than novice users. For example, one of them mentioned, ``\textit{After a summary of possible ways to fix my walking ability to reduce the risk of heart disease, I delved deeper into the examples and kept asking questions.}'' In contrast, novice users rarely asked detailed follow-up questions, often limiting their responses to simple expressions of gratitude or brief agreements with the chatbot’s suggestions (as shown in \Cref{fig:summary_conv_history}). For instance, one of them mentioned: ``\textit{Thank you for these suggestions. I will try to follow these instructions to manage my health better}''. This pattern suggests a potential over-reliance on the agent for novice users \cite{PassiVorvoreanu2022, zhai2024effects}.

(2) \emph{\textbf{ Dialogue-based conversations for informed users}}: Connected to the previous theme, we found out that the novice users were less interested in establishing a dialogue. On the contrary, the informed end users highlighted the benefit of asking follow-up questions to have a proper dialogue-based conversation. For instance, one of them mentioned: ``\textit{Sometimes, when we interact with a real physician, there is less scope to ask follow-up questions to understand their recommendations and instructions in more depth. But now I can ask any number of follow-up questions without the bot judging me}''. This ability to ask follow-up questions is crucial for establishing proper dialogue-based communication between the agent and the user for a higher sense of perceived actionability of the augmented recommendations. Another one remarked: ``\textit{Asking follow-up questions helped me get a detailed answer, with steps}''.

Furthermore, our analysis of the system interaction data revealed that a larger proportion of informed end users (65\%) engaged with the visual explanations to validate the counterfactual recommendations, compared to only about 29\% of novice users. This finding further suggests the possibility of over-reliance from novice users as they were not very keen on validating the augmented recommendations. Surprisingly, very few participants (just 4 out of 34) from either group interacted with the ``\textit{ice-breaker questions}'' to initiate their conversation. This finding raises questions about the relevance of these suggested questions in chatbot applications.  Nevertheless, we did not observe any hallucinated or harmful response from the conversation history data.  


\subsection{How do conversational agents impact the understanding and trust of novice and informed end users? (RQ2)}
\begin{figure}[h]
\centering
\includegraphics[width=0.99\linewidth]{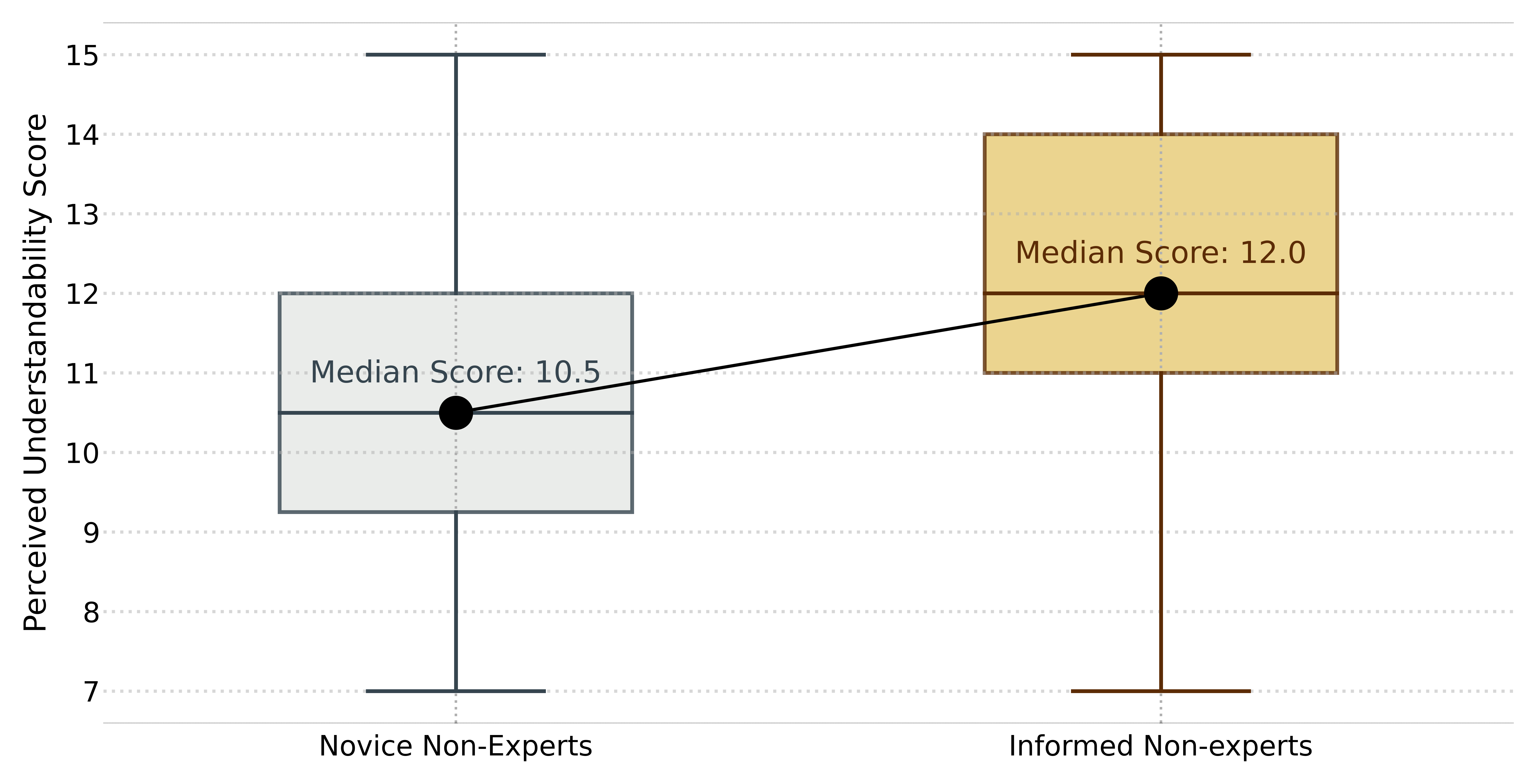}
\caption{Difference in perceived understandability of augmented counterfactuals between novice and informed users.}
\Description[Change in perceived understandability ]{Box plots showing the difference in perceived understandability of augmented recommendations between novice end users and informed end users.}
\label{fig:perceived_understandability}
\end{figure}

The overall perceived understandability of the augmented counterfactuals was rated highly, with an average score of 11.5 out of 15. The scores were particularly higher for the informed users by approximately 11\% on average than the novice users. This difference is visually represented in the box plots in \Cref{fig:perceived_understandability}. Despite observing a clear difference in the scores between these two groups, this difference was not statistically significant using a Mann-Whitney U-test ($U=87.0, p=.064$). Nevertheless, these insights suggest that informed users generally demonstrated a higher level of understanding of the augmented counterfactuals than their counterparts.

However, the difference in perceived trust scores between the two groups was minimal, with novice users showing a marginally higher score by 3\% on average compared to informed users. This difference in the perceived trust scores between the two groups was not statistically significant using Mann-Whitney U-test ($U=154.5, p=.586$). Interestingly, these findings suggest that despite having lower perceived understandability, novice users exhibited marginally higher levels of trust in the system. 

To understand the possible justification for this insight, we delved into the qualitative data captured in our study. Consequently, the following theme was identified that justifies the marginally higher levels of trust for novice users:

(1) \emph{\textbf{Lack of awareness of potential pitfalls of LLMs}}: Since the novice users lacked prior experience with LLM-based chatbots, they were unaware of common issues like hallucination. As a result, they rarely asked follow-up questions to verify the accuracy or the relevancy of the responses. Most simply assessed whether the recommendations to reduce elevated CVD risk seemed reasonable and accepted them if they did. For instance, one participant noted, ``\textit{I trust the application as I did not notice anything that could lead me to mistrust it.}'' Another commented, ``\textit{I feel the chatbot had enough detail and credibility with its answers.}'' In contrast, informed users were more sceptical, often probing the chatbot with detailed follow-up questions, including unrelated queries, to test its responses. One participant explained, ``\textit{I tried to probe the bot by asking about the function of a ballpoint pen to check if it answered unrelated questions.}'' Additionally, the informed group's prior knowledge of LLM limitations led to lower trust: ``\textit{This is a personal feeling about AI chatbots: we’ve all heard about ChatGPT mishaps with medical advice, where it provides a plausible answer that turns out to be completely wrong}''. Therefore, a lack of awareness of the known limitations of LLM chatbots could be a potential reason for novice users' over-reliance and higher levels of trusts.


\subsection{How does perceived taskload differ between novice and informed users when using the chatbot application? (RQ3)}

\begin{figure}[h]
\centering
\includegraphics[width=0.99\linewidth]{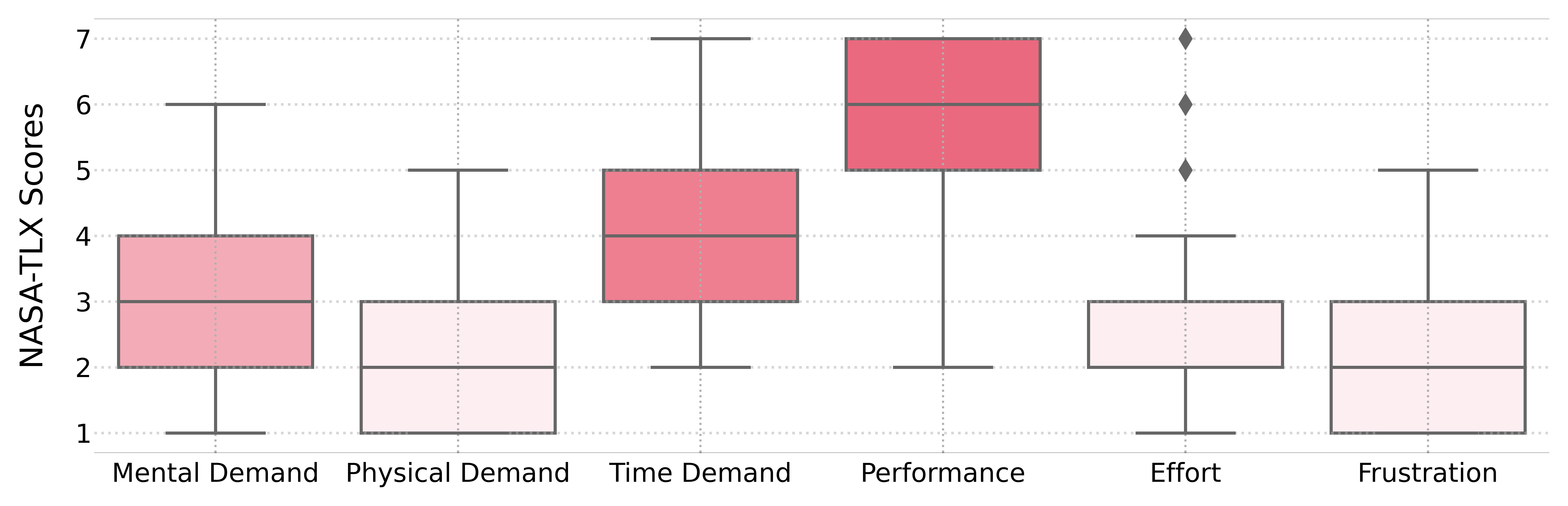}
\caption{Box plots showing the results of perceived taskload assessment using NASA-TLX.}
\Description[NASA-TLX results ]{Box plots showing the results of perceived taskload assessment using NASA-TLX.}
\label{fig:nasa_tlx}
\end{figure}

The results of the NASA-TLX assessment demonstrate low levels of mental demand, physical demand, effort, and frustration when interacting with the application. While time demand was slightly higher, participants rated the system's performance very highly. \Cref{fig:nasa_tlx} presents a summary of these results. These findings suggest that the system enables users to effectively achieve actionable recourse with minimal cognitive and physical strain.

Additionally, using a Mann-Whitney U-test, we found that the difference in the overall perceived taskload between the novice and the informed users was not statistically significant ($U=112.0, p=.33$). This finding indicates that both groups had similar perceived task loads when interacting with the system.

\noindent We analysed the qualitative data to understand participants' high ratings of the system performance, uncovering the following themes:

(1) \emph{\textbf{Easy to follow step-by-step recommendations}}: Participants appreciated how the chatbot provided detailed yet easy to follow, step-by-step recommendations that are practically feasible: ``\textit{The suggestions it gave were backed by steps that you can actually take to help improve your heart's condition}''. They mentioned that the recommendations were very clear and actionable: ``\textit{It does provide each part in an easy to follow way and it breaks down each action}''. These remarks highlight the benefits of augmenting counterfactual instances with a conversational AI agent.

(2) \emph{\textbf{Visual explanations enhanced the understanding of augmented counterfactuals}}: Many participants appreciated the inclusion of the visual explanations through the interactive plots as these helped them understand the recommendations much better by giving an overview of the patient's health conditions. For example, one of them mentioned: ``\textit{The graphs help me keep track of the choices I make based on the chatbot's suggestion. Visualising the current health conditions and the potential changes after following the suggestions makes it much easier to understand than text only}.''


\section{Discussion}

\subsection{Over-reliance of Novice End Users} 

While our study did not directly measure over-reliance on augmented counterfactuals, interactions of novice users and their limited awareness of LLM issues indicate a tendency to over-rely on augmented counterfactuals. Over-reliance on AI arises when users are uncertain about how much to trust it \cite{PassiVorvoreanu2022, zhai2024effects}, often leading to acceptance of incorrect recommendations \cite{PassiVorvoreanu2022}. This observation raises an important question for future research: ``\textit{Do agent-augmented counterfactuals foster over-reliance among users?}'' However, our findings showed no significant signs of over-reliance among informed users, who were more cautious with the chatbot responses. We recommend explicitly informing non-expert users about known issues with LLMs agents, such as hallucinations, through tutorial videos, manuals, or in-app warnings. These steps could help calibrate user reliance on conversational AI for actionable recourse, a topic future research should explore further.

\subsection{Hallucination -- The Elephant in the Room} 
Despite implementing moderation guidelines to minimise hallucinated outputs, recent works increasingly confirm that completely eliminating hallucination is impossible \cite{xu2024hallucinationinevitableinnatelimitation}. While no hallucinated responses were observed in our experiments, we recommend raising broader awareness about hallucination and other common concerns associated with LLMs before users become overly reliant on the generated recommendations. A warning should be provided, advising users to consult trained domain experts (such as healthcare professionals) if any recommendations seem impractical or irrelevant. Additionally, as some informed users attempted to test the agent with off-topic queries, stricter guardrails should be implemented to prevent hallucinations triggered by inputs beyond the chatbot’s intended scope.

\subsection{Importance of Ice-Breaker Questions} 
Surprisingly, while our exploratory study highlighted the need for \textit{ice-breaker questions} to initiate conversations with the chatbot, interaction data from the final user study revealed minimal engagement with these questions. This observation contradicted findings from previous studies, which had recommended the inclusion of such suggested questions \cite{Slack2023, freed2021conversational}. Although multiple factors could contribute to this observation, we believe that the specific task-oriented nature of the application minimised the need for suggested ice-breaker questions. It is likely that such questions are more valuable for general-purpose chatbots designed for broader, more diverse use cases. 

\subsection{Domain Knowledge Inclusion Through RAG}
Given the increasing advocacy for involving domain experts in the development and fine-tuning of prediction models \cite{bhattacharya2024exmos, kulesza_principles_2015, teso_leveraging_2022}, there is a compelling opportunity to extend their involvement for achieving actionable recourse, ensuring more accurate and context-sensitive outcomes. One method for incorporating domain experts' prior knowledge is through the Retrieval Augmented Generation (RAG)  \cite{gao2024retrievalaugmentedgenerationlargelanguage}. RAG creates a knowledge base that allows the conversational agent to retrieve up-to-date information and guidelines from trusted sources. For instance, in the medical domain, this knowledge base could include the latest guidelines from reputable organisations such as the WHO and the CDC. By using RAG, the agent can ensure that its responses are both current and accurate, significantly reducing the likelihood of hallucinations affecting the agent-augmented counterfactuals.

\subsection{Broader Applicability of Our Guidelines}
To increase the broader applicability of our guidelines,  we suggest embedding augmented counterfactual recommendations within pre-consultation chatbots \cite{BrennaPreConsultationCHI2024}. This initial step facilitates crucial information sharing between experts and users, promoting user-centred services and alleviating expert workload. Providing personalised, actionable insights via augmented counterfactuals empowers users to understand their data and explore potential outcomes. Consequently, users receive tailored advice on how to achieve desired results, which streamlines consultations, allows experts to concentrate on intricate matters, and enables users to confidently investigate "what-if" scenarios before engaging with experts. Recognising the strength of our current guidelines, we emphasise the need for more in-depth user studies across diverse contexts to further refine and generalise their effectiveness. This iterative process will ensure our framework remains adaptable and impactful in the realm of user-centred agent-augmented counterfactual explanations.

\subsection{Limitations}
Despite all precautionary measures, we could not avoid the following limitations during this research:

\noindent (1) \textit{Sample Size and Diversity of User Study Participants}: While we made every effort to recruit participants with diverse demographic backgrounds, our participants belonged only to a limited set of geographical regions. Future studies should address this limitation by increasing the sample size and recruiting participants with more diverse demographic backgrounds. Additionally, with a sufficiently large participant pool, a between-subject study comparing augmented and non-augmented counterfactuals would offer a more comprehensive understanding of their trade-offs.

\noindent (2) \textit{Constraints with the front-end framework}: The front end of the chatbot application was limited by the capabilities of Streamlit. While Streamlit facilitated rapid development, it has known constraints in terms of memory and computational power, making it less suitable for applications with a high volume of concurrent users. Additionally, the large size of our dataset sometimes resulted in slow model retraining and counterfactual generation. 

\noindent(3) \textit{More robust evaluation measures}: While exploring various evaluation methods from prior research, we identified opportunities for more robust metrics. For example, Singh et al.'s validated toolbox for measuring actionability \cite{singh2024actionabilityassessmenttoolexplainable} could enhance the assessment of perceived actionability. Similarly, objective understandability evaluations, as used by Bhattacharya et al. \cite{bhattacharya2024exmos}, might complement Hoffman et al.'s subjective questions \cite{hoffman2019metrics}. Additionally, studies suggest that assessing trust requires longitudinal evaluations to account for its gradual development \cite{zafari2024trust}. Future work should incorporate these measures for a more comprehensive evaluation.

\section{Conclusion}
This paper introduces our general guideline for augmenting counterfactual explanations with conversational AI agents tailored for non-expert. Using these guidelines, we developed a healthcare chatbot that offers actionable recommendations to patients at elevated CVD risk. Through an extensive mixed-methods study with 34 participants we found that users with prior experience in conversational agents engaged more effectively with augmented counterfactuals, while novice users showed potential over-reliance. Based on these findings, we offer recommendations for designing efficient chatbots that deliver effective and actionable insights to users using agent-augmented counterfactuals.

\begin{acks}
We acknowledge the helpful feedback from Maxwell Szymanski, Yizhe Zhang and Grzegorz Meller that improved this research work. We also extend our gratitude to the participants of our user study. The funding support for this research was obtained from DSTRESS (HBC.2024.0681), Research Foundation–Flanders (FWO grants G0A4923N and G067721N),  Flanders AI Research Program (FAIR) \cite{BhattacharyaCHI2025, UmapBhattacharya2024} and KU Leuven internal funds (C14/21/072) \cite{BhattacharyaCHIDC, Bhattacharya2024HowGoodIsYourExplanation}.
\end{acks}

\bibliographystyle{ACM-Reference-Format}
\bibliography{references}

\end{document}